\documentclass[fleqn,usenatbib]{mnras}

\usepackage{newtxtext,newtxmath}

\usepackage[T1]{fontenc}
\usepackage{ae,aecompl}

\usepackage{graphicx}	
\usepackage{amsmath}	
\usepackage{amssymb}	
\usepackage{lipsum}
\usepackage{rotating}
\usepackage{pdflscape}
\usepackage[usenames,dvipsnames,svgnames,x11names]{xcolor}
\usepackage{ulem}

\newcommand{\code}[1]{\texttt{#1}}

\title[The Mergers in A2256]{The Mergers in Abell 2256: Displaced Gas and its Connection to the Radio-emitting Plasma}

\author[J. P. Breuer et al.]{J. P. Breuer$^{1,2}$\thanks{E-mail: jeanpaul.breuer@gmail.com},
N. Werner$^{3,1,4}$,
F. Mernier$^{3,5,6,7}$,
T. Mroczkowski$^{2}$,\newauthor
A. Simionescu$^{7,8,9}$,
T. E. Clarke$^{10}$,
J. A. ZuHone$^{11}$,
L. Di Mascolo$^{12}$
\\
$^{1}$Department of Theoretical Physics and Astrophysics, Masaryk University. Kotl{\'a}{\v r}sk{\'a} 2, Brno, 611 37, Czech Republic\\
$^{2}$European Southern Observatory, Karl-Schwarzschild-Stra{\ss}e 2, 85748 Garching bei M{\"u}nchen, Germany\\
$^{3}$MTA-E{\"o}tv{\"o}s University Lend{\"u}let Hot Universe Research Group, P{\'a}zm{\'a}ny P{\'e}ter s{\'e}t{\'a}ny 1/A, Budapest, 1117, Hungary\\
$^{4}$School of Science, Hiroshima University, 1-3-1 Kagamiyama, Higashi-Hiroshima, Hiroshima 739-8526, Japan\\
$^{5}$Institute of Physics, E\"otv\"os University, P\'azm\'any P\'eter s\'et\'any 1/A, Budapest, 1117, Hungary\\
$^{6}$ESA/ESTEC, Keplerlaan 1, 2201 AZ Noordwijk, The Netherlands\\
$^{7}$SRON Netherlands Institute for Space Research, Sorbonnelaan 2, 3584 CA Utrecht, The Netherlands\\
$^{8}$ Leiden Observatory, Leiden University, PO Box 9513, 2300 RA Leiden, The Netherlands\\
$^{9}$ Kavli Institute for the Physics and Mathematics of the Universe (WPI), The University of Tokyo, Kashiwa, Chiba 277-8583, Japan \\
$^{10}$Naval Research Laboratory, 4555 Overlook Avenue SW, Code 7213, Washington, DC 20375, USA\\
$^{11}$Center for Astrophysics | Harvard \& Smithsonian, 60 Garden St., MS-67, Cambridge, MA 02138, USA\\
$^{12}$Max-Planck-Institut für Astrophysik, Karl-Schwarzschild-Strasse 1, Garching D-85741, Germany
}

\date{Accepted XXX. Received YYY; in original form ZZZ}

\pubyear{2020}

\begin{document}
\label{firstpage}
\pagerange{\pageref{firstpage}--\pageref{lastpage}}
\maketitle

\begin{abstract}
We present the results of deep {\it Chandra} and {\it XMM-Newton} X-ray imaging and spatially-resolved spectroscopy of Abell 2256, a nearby ($z=0.058$) galaxy cluster experiencing multiple mergers and displaying a rich radio morphology dominated by a large relic. The X-ray data reveals three subclusters: (i) the `main cluster'; (ii) the remnant of an older merger in the east of the cluster with a $\sim 600$~kpc long tail; (iii) a bright, bullet-like, low-entropy infalling system, with a large line-of-sight velocity component. The low-entropy system displays a 250~kpc long cold front with a break and an intriguing surface brightness decrement. Interestingly, the infalling gas is not co-spatial with bright galaxies and the radio loud brightest cluster galaxy of the infalling group appears dissociated from the low entropy plasma by $\sim 50$~kpc in projection, to the south of the eastern edge of the cold front. Assuming that the dark matter follows the galaxy distribution, we predict that it is also significantly offset from the low-entropy gas. Part of the low frequency radio emission near the cold front might be revived by magnetic field amplification due to differential gas motions. 
Using analytical models and numerical simulations, we investigate the possibility that the supersonic infall of the subcluster generates a large scale shock along our line-of-sight, which can be detected in the X-ray temperature map but is not associated with any clear features in the surface brightness distribution.
\end{abstract}




\section{Introduction}
In the $\Lambda$CDM model of hierarchical structure formation, clusters of galaxies are the largest gravitationally bound objects to form, serving as probes into the history of large scale structure and galaxy formation, and functioning as laboratories for studying both plasma physics and the underlying cosmology. These systems are driven by the gravity of dark matter, and grow over time through violent mergers with other clusters and via constant accretion through the filamentary strands of the cosmic web. Tremendous energy from merging events is released in the form of shocks, which propagate through the intracluster medium (ICM) and heat the gas up to temperatures of $10^{7}-10^{8}$ K, shining brightly in the X-rays. Moreover, in major mergers, the collisionless dark matter distribution is sometimes found to be dissociated from most of the baryonic component \citep{Clowe2006}. Since the majority of the baryonic matter in clusters is the hot ICM, X-ray observations serve as a powerful probe in investigating the characteristic properties inherent in the formation of large scale structures \citep{Sarazin86,HansWerner}.  


The characteristic temperature, density, and chemical abundance of infalling gas from a subsystem naturally differs from the gas of the main cluster, thus it is possible to observe many variations of instabilities and substructures which develop in the ICM during mergers as the various plasma populations mix, ranging from sub-kpc to Mpc scales. 
The most frequently observed large scale X-ray substructures in merging systems are cold fronts, which can be subdivided into `merger-remnant' and `sloshing' cold fronts, as well as shock fronts \citep[see][for reviews]{zuhone_roediger_2016,Maxim07}. 
Merger-remnant cold fronts can be identified as a contact discontinuity showing a temperature and density jump between a near continuous pressure profile, and are formed as a result of a cool, subsonically-moving subhalo propagating through a hotter ambient gas. In a similar way, sloshing cold fronts can be seen in the case when the dense gas associated with a cool core is displaced within the main gravitational potential well, showing a spiral pattern or concentric arcs, depending on the line of sight orientation with respect to the merger plane \citep{Ascasibar06,Roediger11}. Shock fronts are driven when a subcluster moves supersonically through the ICM, producing sharp jumps of density, temperature, pressure, and entropy across the boundary.

Various types of shocks are formed in the ICM during dynamical events: {\it external accretion shocks} around clusters from accreting gas, {\it internal} and {\it turbulent shocks} within clusters from infalling clumps, as well as {\it axial} and {\it equatorial merger shocks} from larger merger events \citep{Ha2018, Gu19}. Axial shocks are more easily detected than equatorial merger shocks as they occur after core passage between two systems, for example, in the case of the well-studied Bullet Cluster \citep{Markevitch2002, Markevitch2006, Clowe2006, Bradac2006, Paraficz2016, Luca2019}. On the other hand, equatorial shocks are more difficult to observe, as they are triggered only when systems are in a pre-merger state, before two subclusters have actually merged, a fact that also makes them very different from the axial shocks with regards to velocities, distances, and time scales.

A textbook example of complex merging galaxy clusters is Abell 2256. This is a nearby ($z = 0.0581$), rich (richness class 2) system with strong X-ray emission ($L_{\rm X} \sim 10^{45}$ erg s$^{-1}$) whose merging scenario has been extensively debated in the literature. Previous observations with \textit{ROSAT} by \citet{Briel91} and \textit{ASCA} by \citet{Maxim96} revealed two X-ray surface brightness peaks near the center. One of the peaks is close to the geometric center of the cluster, while the other is around 250 kpc to the west and is considered to be a merging subcluster with a lower temperature than the main cluster \citep{Maxim96}. An analysis of \textit{Chandra} observations by \citet{Sun02} revealed a third subgroup to the east, nicknamed the ``shoulder''. An associated cold front was found using \textit{XMM-Newton} observations by \citet{Bourdin08}, which showed a hot bow-like temperature structure. 

\begin{figure*}
	\centering
	\includegraphics[width=0.8\textwidth]{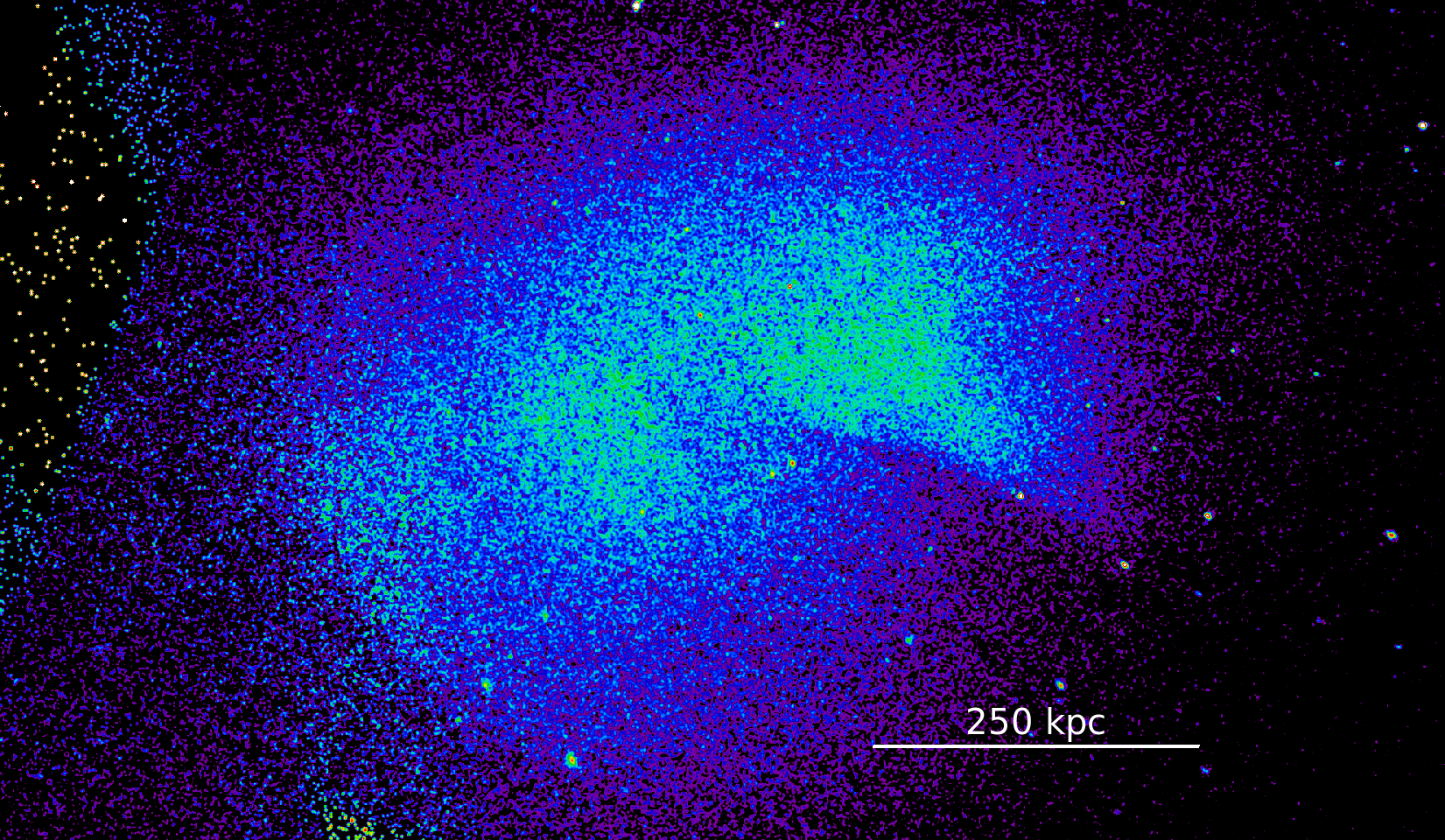}
	\includegraphics[width=0.8\textwidth]{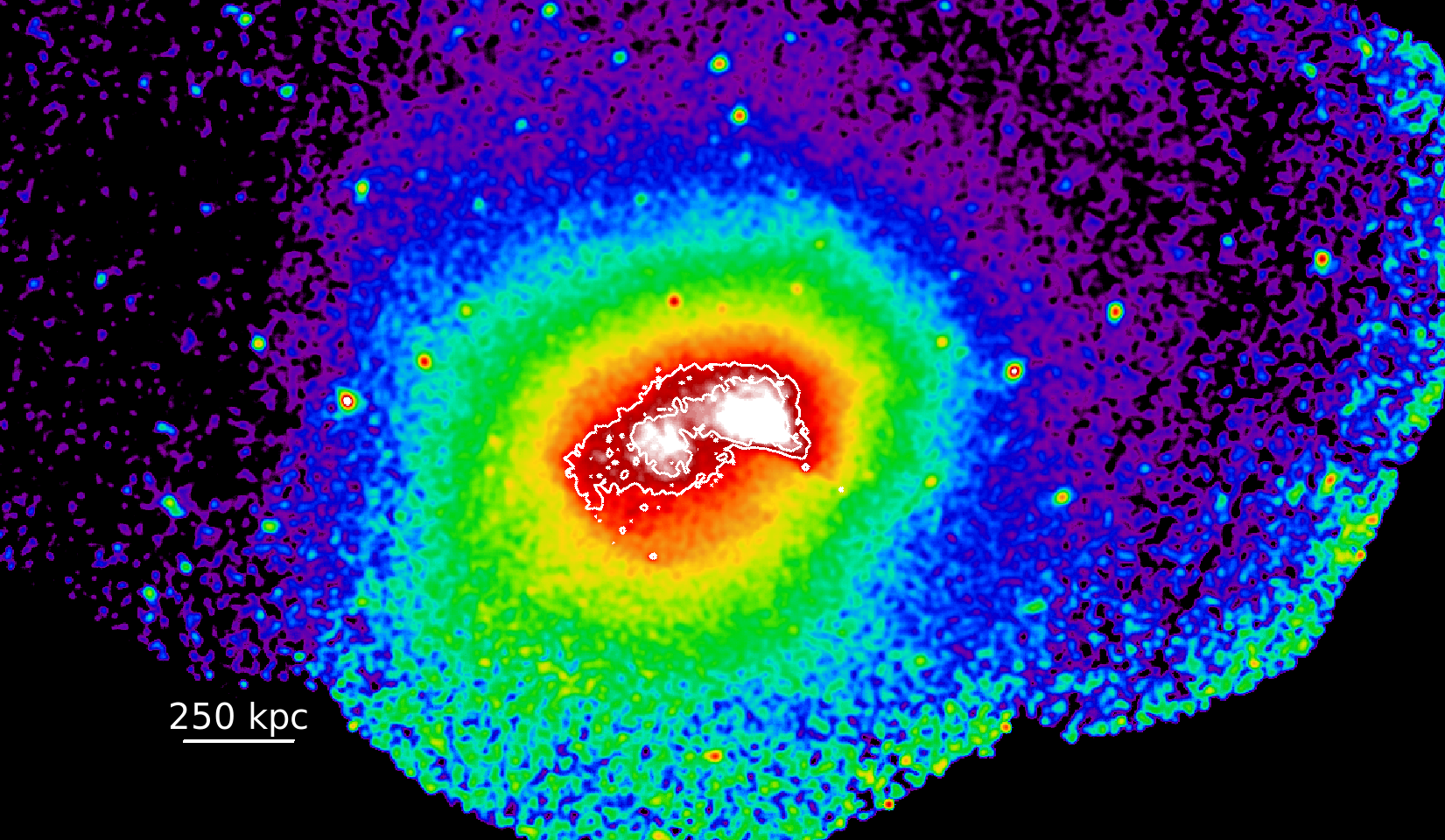}
    \caption{{\it Top:} An exposure corrected {\it Chandra} broad band (0.5--7~keV) image of Abell 2256 using all available observations. North is up and east is left (for this and all subsequent images). {\it Bottom:} An exposure corrected {\it XMM-Newton} soft band (0.3--2 keV) image of Abell 2256 overlaid with white {\it Chandra} contours. The enhancement of surface brightness at the edges of the field of view is due to flaring events.}
    \label{fig:xray-raw}
\end{figure*}

Historically, there have been many studies on Abell 2256, with the first publications dating back now almost 50 years \citep{First}. The scientific interest in Abell 2256 has been mostly dominated by radio observations due to the exceptional radio features, although there have also been various optical and X-ray follow up studies of the cluster. The velocity dispersion of the galaxies was measured by \citet{Faber77} to have, at the time, one of the highest velocity dispersions known in galaxy clusters ($\sim$ 1700 km s$^{-1}$). \citet{Fabricant89} increased the sample from 14 to 89 galaxies and suggested that this system is a merger in progress. A follow-up study by \citet{Berrington02} again increased the sample of radial velocities to 319 galaxies, 277 of which are cluster members, and added photometry for 861 galaxies, showing strong evidence for a merger event and also presenting new evidence for an additional group of galaxies inside the relic region to the north of the main cluster, which is moving south to eventually merge. Recent observations with \textit{Suzaku} by \citet{Tamura11} also suggested gas bulk motion within Abell 2256 using the Doppler shift of the Fe-K X-ray line emissions, to show that the velocity difference between the main and the western sub-component of the cluster is approximately 1500 km s$^{-1}$.

Extensive discussion of the various radio sources seen in Abell 2256 and their spectral properties is also provided by \citet{Intema09} and \citet{Kale10}. The 1.2 Mpc radio halo component at low frequencies were discussed in \citet{Weeren12,Brentjens08,Tracy06}, while \citet{Ozawa15} show a polarimetry study. For greater context on the implications of the relics on particle acceleration see \citet{Weeren19} and \citet{Trasatti15}. 

Using the most complete \textit{Chandra} and \textit{XMM-Newton} archival observations of Abell 2256 to date, taking advantage of deeper and more detailed X-ray imaging and spectroscopy than seen in previous work \citep[e.g.][]{Sun02,Briel91,Maxim96,Fusco2000,Bourdin08}, this paper investigates the gas morphology and detailed substructure of the cluster. The rich, complex X-ray morphology is compared with publicly available radio observations of this system. Section~\ref{sec:observations} explains the methods used in the data reduction and analysis. Section~\ref{sec:results} describes the main results and Section~\ref{sec:discussion} discusses these results. 
Throughout the paper, we assume the standard cosmology with $\Omega_m = 0.286$, $\Omega_\Lambda = 0.714$, and $H_0 = 69.6$. Consequently, at $z = 0.0581$, 1 arcmin corresponds to 67.9 kpc. Unless stated otherwise, the error bars correspond to 68\% of confidence interval.

\section{Observations and Data Reduction}\label{sec:observations}

\subsection{\textit{Chandra} Observations}\label{sec:chandra}
The data from five \textit{Chandra} pointings (see Table~\ref{tab:observations}) were reprocessed using \texttt{repro} from the level-1 event lists with the standard software packages using the most recent versions of CIAO and CALDB (versions 4.11 and 4.8.2, respectively). The good time interval of the observations after removing bad pixels and filtering periods of flaring 
are also summarized in Table~\ref{tab:observations}. Totalling all available {\it Chandra} observations gives approximately 215 ks of observing time; however, only 186 ks were used for creating the data products due to the remaining exposures having experienced flaring contamination which could bias the spectral analysis. All observations were used for creating the surface brightness image, but only observations 2419, 16129, 16514, 16515, and 16516 were considered in further spectral analysis. Blank-sky background files were processed in a manner similar to the source observations and were scaled by the data over background count ratio in the 9.5-12 keV energy range.

\begin{figure*}
	\includegraphics[width=\textwidth]{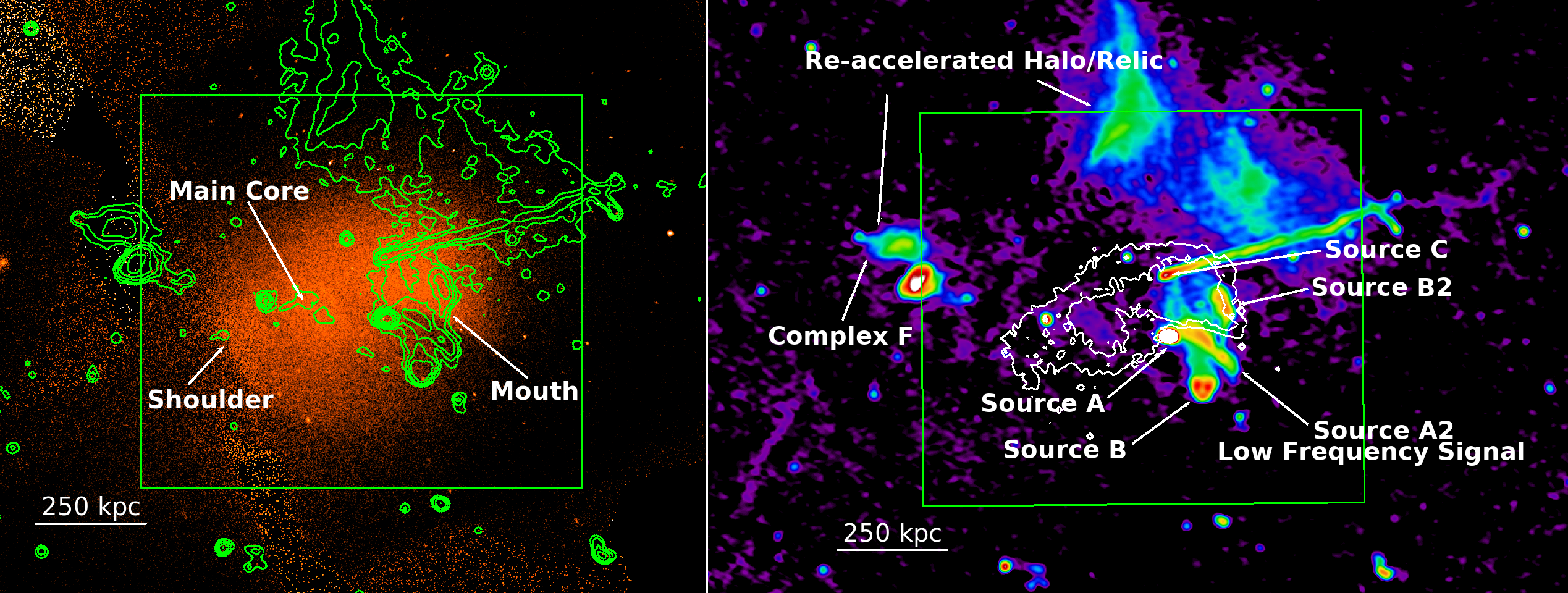}
    \caption{A \textit{Chandra} broad band X-ray image of Abell 2256 (0.5--7~keV) with the 325 MHz GMRT radio contours in green. (\textit{left}) A 325 MHz \textit{GMRT} image with the Chandra X-ray contours in white (\textit{right}). The box indicates the reduced field-of-view used for the subsequent image and spectral analysis. BCG1 is located in the main core, while BCG2 is associated with radio source A.}
    \label{fig:xray-radio}
\end{figure*}

\vspace{0.25cm}
\subsubsection{Image Processing}\label{sec:imageanal}
Fig.~\ref{fig:xray-raw} shows the background subtracted and exposure corrected \textit{Chandra} image in the 0.5 to 7 keV energy range. 
The left panel of Fig.~\ref{fig:xray-radio} shows another \textit{Chandra} image overlaid with radio contours from the 325 MHz GMRT observation. The right panel of Fig.~\ref{fig:xray-radio} shows the 325 MHz GMRT image of the same field with overlaid {\it Chandra} X-ray contours. The 325 MHz radio image in Fig.~\ref{fig:xray-radio}, referenced throughout this paper, is the same dataset used in the work of \citet{Intema09}. The box in both images corresponds to the region where the further X-ray data analysis was performed.
Point sources were initially detected and excluded from subsequent analysis using the \texttt{wavdetect} tool using scales of 1, 2, 4, 8, 16, and 32 pixels; however, visual inspection was required to remove additional point sources from the image.

Fig.~\ref{fig:ggm} shows the distribution of sharp features in the X-ray image, obtained using the directional Gaussian gradient magnitude (GGM) filter \citep[see][]{sanders2016a,sanders2016b,Walker16}. To further emphasize the substructures, \texttt{SHERPA} was used to create a relative deviation image of the \textit{Chandra} data with respect to the best fitting 2D elliptical beta model (\texttt{beta2d}). C-statistics and a Monte Carlo method were used for finding the best fitting parameters for the center position (\texttt{xpos, ypos}), ellipticity (\texttt{ellip}), and roll angle (\texttt{theta}). The best fitting position for the cluster center was determined to be at approximately (17:03:56.2853 +78:38:30.156), the ellipticity is 0.30331, with a corresponding roll angle of 0.413305 radians. Fig.~\ref{fig:beta} shows the relative deviation image after a radius = 14.0, sigma = 7.0 Gaussian smoothing kernel was applied. The center of the best fitting centroid is marked with an `X'.

The edges found in the GGM and beta model subtracted images were further explored using \texttt{Proffit} v.1.5 \citep{Proffit}. We fit the surface brightness profiles by projecting a spherically-symmetric, discontinuous, broken power-law density distribution \citep{owers2009}.  The free parameters of our model are the normalization, the inner and outer density slopes, and the amplitude and radius of the density jump. The results are presented in Sect.~\ref{sec:results}.

\vspace{0.25cm}
\subsubsection{Thermodynamic Maps\label{sec:specanal}}
The thermodynamic maps for Abell 2256 were computed to study the internal structure of the ICM; specifically, we use the contour binning algorithm from \citet{Sanders06}, which creates regions with a pre-selected signal-to-noise (S/N) ratio, while conserving the general contours of the data by grouping neighboring pixels of similar surface brightness, yielding statistically independent regions. 

Spectra were extracted from each region determined by the contour binning algorithm, and were fit using \texttt{XSPEC} (version 12.10.0c and AtomDB version 3.0.9). The model for a plasma in collisional ionisation equilibrium (\texttt{apec}) with photoelectric absorption (\texttt{phabs}) was used, with fixed values for the metallicity (0.3 Solar), redshift (z = $0.0581$), and neutral hydrogen column density ($N_\text{H}$ = $4.1 \cdot 10^{20}$ cm$^{-2}$), as determined by the LAB (Leiden/Argentine/Bonn) radio HI survey \citep{Kalberla05}. Unless otherwise mentioned, this spectral fitting methodology applies to any other spectral analysis considered for the rest of the paper.

Fig.~\ref{fig:jointmaps_black} shows maps of the temperature, density, pressure, and entropy. The temperature \textit{kT} was derived directly from the spectral modeling, while the electron density \textit{$n_\textrm{e}$} was computed from a normalization parameter, {\it norm}, formally defined as
\begin{equation}
\textit{norm} = \frac{10^{-14}}{4\pi (d_\textsc{a} (z+1))^2}\int n_\textrm{e} n_\textrm{p} dV \qquad,
\label{eq:norm}
\end{equation}
where $d_\textsc{a}$ is the angular size distance, $z$ is the redshift, and $\int n_\textrm{e} n_\textrm{p}$ is the integrated cluster emissivity over the chosen volume of gas in the cluster. The electron density $n_\textrm{e}$ and ion density $n_\textrm{p}$ are related as $n_\textrm{e} = 1.18 n_\textrm{p}$. The electron density is calculated under the simplified assumption of a constant line-of-sight depth of 1 Mpc. The gas pressure and entropy are then determined as $P = n_\textrm{e} kT$ and $K = kT n_\textrm{e}^{-2/3}$, respectively. 

All maps were computed with S/N values of 33 and 70 (1000 and 5000 counts per region, respectively), however, shown in Fig.~\ref{fig:jointmaps_black} is only the S/N 70 temperature map and S/N 33 density map, along with their combined S/N pressure and entropy maps, all of which had relative errors and propagated uncertainties lower than 10\%. Combining the S/N 70 and 33 region maps yields a better resolved gradient between map regions at the expense of having completely statistically independent bins. This is a decent method in improving the map resolution without suffering from over-smoothing as seen in other binning methods. Results were later double checked with maps of S/N values of 100 and 142 (10,000 and 20,000 counts, respectively), and the conclusions were not altered by the improved statistics from using larger bins.

\subsection{\textit{XMM-Newton}}
Complementary to the \textit{Chandra} data, we also retrieved archival \textit{XMM-Newton} observations of Abell 2256, as listed in Table~\ref{tab:observations}. Eight pointings showing sufficient data quality were downloaded and reduced using the XMM Science Analysis System (SAS) software (v17.0.0). For each pointing, the EPIC MOS\,1, MOS\,2, and pn raw data were obtained using the standard pipeline tasks \texttt{emproc} and \texttt{epproc}. Regrettably, most of the observations were heavily contaminated by flares, and a significant fraction of the exposures had to be filtered out. This was done by grouping each exposure of each detector into regular time bins of 10 and 100 s in the 0.3--10 keV and 10 to 12 keV bands, respectively, then building histograms of their count rates and discarding count rates exceeding 2$\sigma$ from the mean of the best-fit Poisson distribution \citep[for more details, see e.g.][]{Mernier15}. Good time intervals are indicated in the last column of Table~\ref{tab:observations}. In some cases, however, residual, quiescent soft-protons contaminated a given detector and increased dramatically the noise level of its raw image. For this reason, and after visual inspection, we discarded the EPIC MOS data from ObsID: 0112950901, as well as the EPIC pn data from ObsID: 0112500201 and ObsID: 0401610101.

Images were extracted for each detector of each observation using the SAS task \texttt{evselect}, then background-subtracted using appropriate filter wheel closed observations. In addition, exposure maps were obtained for each image using the task \texttt{eexpmap}. In order to minimise the effect of loss of information from detector chip gaps, background-subtracted images and exposure maps were first merged separately before normalising the merged raw image by the merged exposure map. The final result is shown for the 0.3 to 2 keV band in the bottom panel of Fig.~\ref{fig:xray-raw}. Although the inner $\lesssim$500 kpc region of the cluster is well mapped by {\it XMM-Newton}, the anomalously bright edges of the detectors suggest an imperfect background subtraction and correction by the exposure maps, due to a significant contamination of regions of lower surface brightness by residual soft protons. We have specifically checked that this effect remains when restraining the filtered count rates to 1$\sigma$ from the mean of the distribution, as well as when excising the outer 10 arcmin of each individual pointing.

In principle, the \textit{XMM-Newton} data could be used to explore surface brightness features associated with the hot regions; however, the pervasive soft-proton contamination would interfere with any quantitative interpretation. This is because the soft-protons have a different vignetting curve than X-ray photons (which is different still from the instrumental background that is not vignetted at all) and the strong soft proton contamination could result in extraneous spatial/radial trends in the surface brightness profile. In addition to their spatial variation, the protons behave differently in terms of flux and spectral slopes with time, so any attempt to model them in the \textit{XMM-Newton} data might introduce extra biases and uncertainties to derived values. Given the contamination of the \textit{XMM-Newton} data, in the rest of the paper we focus on the \textit{Chandra} observations, allowing to not only derive thermodynamic maps, but also explore surface brightness features while keeping systematic uncertainties under control. Nevertheless, the \textit{XMM-Newton} EPIC image confirms the large scale morphology of the system seen with \textit{Chandra}.







\begin{table}
	\centering
	\caption{Summary of the data, excluding overly contaminated observations. Net exposure time is the good time interval after cleaning the data. $^{(M)}$ EPIC MOS only. $^{(p)}$ EPIC pn only.}
	\label{tab:observations}
	\begin{tabular}{lccr}
	  \hline 
		Telescope & ObsID & Date & GTI (ks)\\
        \hline 
        \textit{Chandra} & 965 & 1999-10-14 & 11.2\\		
        \textit{Chandra} & 1386 & 1999-10-13 & 12.5\\
		\textit{Chandra} & 2419 & 2002-02-14 & 10.3\\		
		\textit{Chandra} & 16129 & 2015-09-28 & 44.5\\
		\textit{Chandra} & 16514 & 2015-09-28 & 44.2\\
		\textit{Chandra} & 16515 & 2015-09-28 & 42.9\\
		\textit{Chandra} & 16516 & 2015-09-28 & 44.2\\
        \textit{XMM-Newton}$^{(M)}$ & 0112500201 & 2002-03-20 & 11.2\\
        \textit{XMM-Newton} & 0112950801 & 2002-04-17 & 5.1\\        
        \textit{XMM-Newton}$^{(p)}$ & 0112950901 & 2002-04-23 & 10.0\\
        \textit{XMM-Newton} & 0112951501 & 2002-06-02 & 8.3\\
        \textit{XMM-Newton} & 0112951601 & 2002-09-22 & 10.1\\
        \textit{XMM-Newton} & 0141380101 & 2003-04-27 & 9.6\\
        \textit{XMM-Newton} & 0141380201 & 2003-06-30 & 11.9\\
        \textit{XMM-Newton}$^{(M)}$ & 0401610101 & 2006-08-04 & 45.0\\
        \hline 
	\end{tabular}
\end{table}

\section{Results}\label{sec:results}

\subsection{General X-ray, Radio, and Optical Properties}\label{sect:results_morphologies}
The X-ray images obtained by {\it Chandra} and {\it XMM-Newton} shown in Fig.~\ref{fig:xray-raw} reveal three distinct regions of increased surface brightness, indicating the presence of at least three sub-clusters. 
The image also shows two surface brightness discontinuities, one in the east and another prominent quasi-linear edge associated with the western subcluster. 
Following the previous nomenclature coined by \citet{Sun02}, the feature observed in the east will continue to be referred to as the ``shoulder''; while henceforth, the central gas distribution will be referred to as either the ``main core'' or the ``body'', and the western merging structure will be called the ``mouth''. The naming convention for the subclusters was chosen due to the general morphology and the sharp and elongated contact discontinuity associated with the western cold front, which is also reminiscent of a certain hungry yellow protagonist from a popular maze arcade game from the 1980's (see Fig.~\ref{fig:xray-raw} top).


The radio observation used in this paper is a public 325 MHz {\it GMRT} dataset as used in \citet{Intema09}, and shows a very different morphology (see Fig.~\ref{fig:xray-radio} right panel). The most prominent feature is the radio relic to the north, which extends about 330 kpc (6.3 arcmin) in the north-south and approximately 588 kpc (11.1 arcmin) in the east-west direction. There are also two head tail radio galaxies, one with an approximate projected length of 445 kpc (8.4 arcmin) which extends in the east-west direction and a smaller one to the west extending in the south-east direction with a projected length of 106 kpc (2 arcmin). To the south, one sees a wide-tail lobe type AGN ({\it Source B}), a radio galaxy ({\it Source A}) and an associated low frequency signal which is only visible under 1 GHz \citep[{\it Source A2}, see][]{Intema09}. \citet{Intema09} propose that A2 is independent from the wide-angle tail AGN located to the south and is instead connected with the radio galaxy to the west.

The optical data presented in \citet{Berrington02} reveal a main galaxy concentration with a BCG associated with the central gas density excess (`the body') and another merging subgroup whose BCG is associated with radio source A. Surprisingly, there are no galaxies spatially associated with the infalling, X-ray bright, low entropy gas (`the mouth') and the BCG of the merging subgroup is displaced by $\sim50$~kpc in projection to the south of the eastern edge of the cold-front. This will be further discussed in Sect.~\ref{sect:discussion_CF}.


\subsection{Detailed X-ray Image Analysis}
\begin{figure}
\includegraphics[width=\columnwidth]{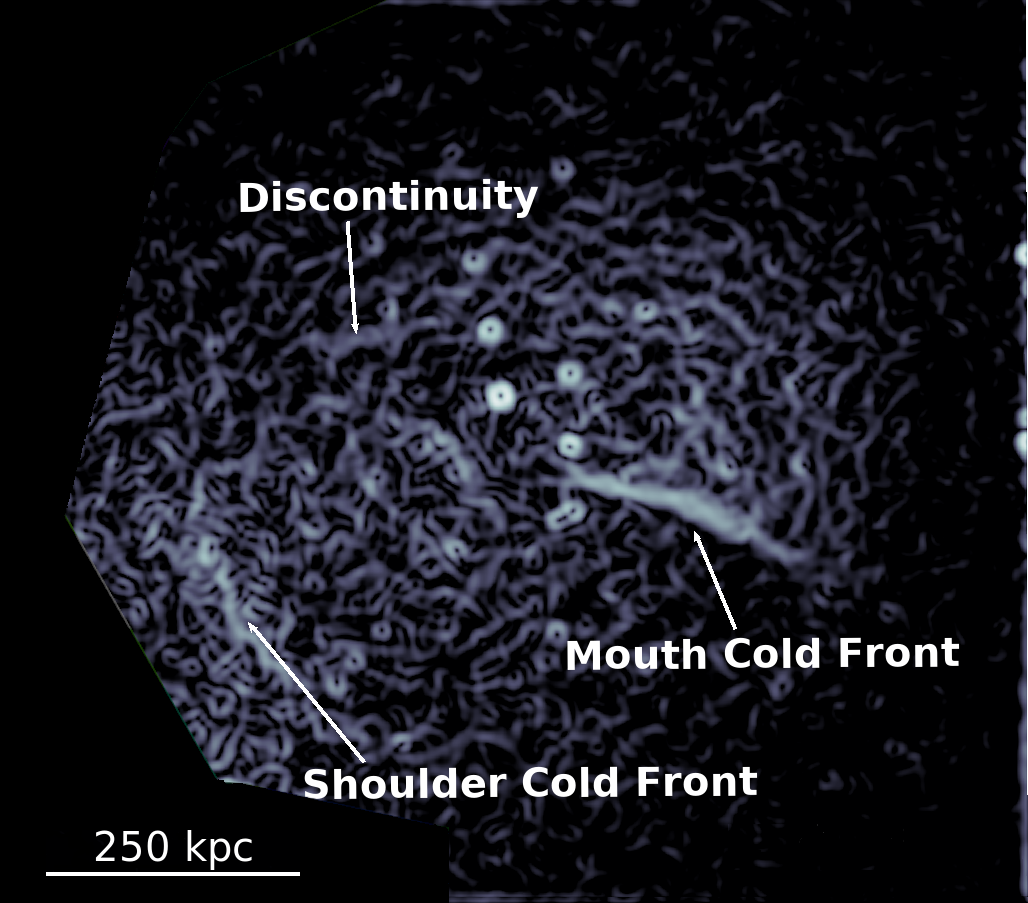}
	\caption{A Gaussian gradient magnitude filter (Sobel kernel) applied to the \textit{Chandra} image to emphasize the surface brightness edges. Of note, is the long, $\sim$250 kpc linear edge at the ``mouth" to the west, a slightly weaker $\sim$200 kpc curved edge to the east, and a third even weaker curve near the center of the cluster, which might be due to gas compression from the merger event. Point sources dotting the center of the image should be ignored and neglected when considering this image.}
\label{fig:ggm}
\end{figure}
\begin{figure}
	\includegraphics[width=\columnwidth]{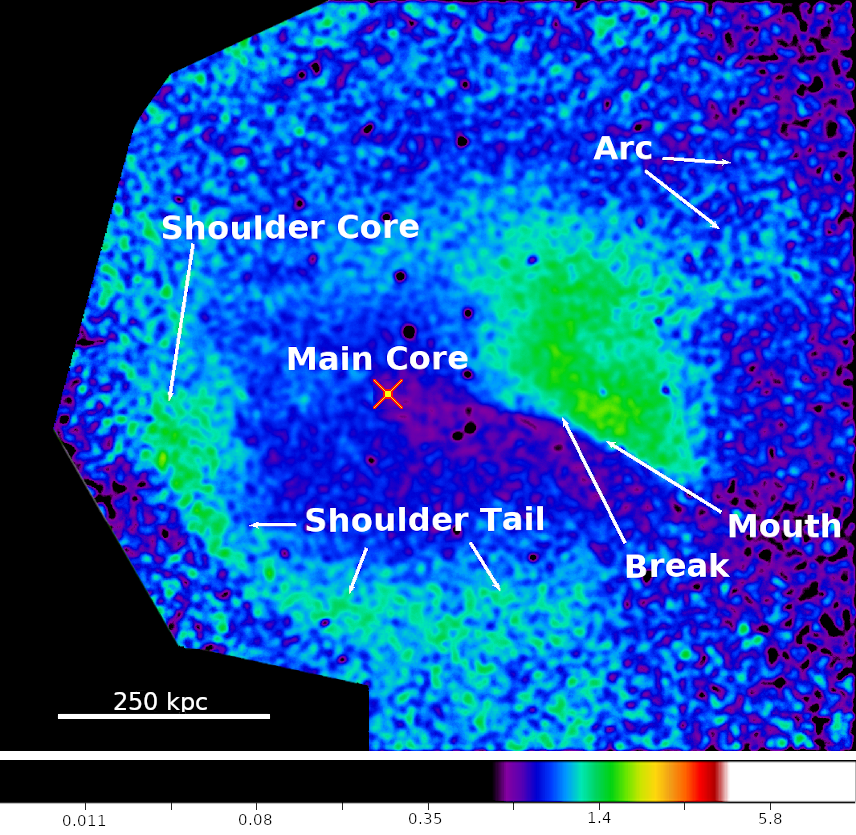}
	\caption{Residuals of the best fit elliptical 2D beta model. After removing the main cluster component, the remaining residuals reveal two subclusters at different stages of a merger. The `X' on the image marks the center of the best fit centroid.}
	\label{fig:beta}
\end{figure}

An edge detection analysis (Fig.~\ref{fig:ggm}) shows strong surface brightness gradients associated with the ``shoulder'' and the ``mouth''. A possible fainter and weaker edge can also be seen in the cluster center. Constant pressure profiles across the two prominent surface brightness discontinuities are in line with previous studies which have shown that they are cold fronts \citep{Sun02,Bourdin08}. While the cold front associated with the ``mouth''  appears remarkably sharp, it also shows a clear break around its centre (see Figs.~\ref{fig:ggm}, \ref{fig:beta}, and \ref{fig:coldfront}).


The result of removing the main cluster component from the image using a 2D beta model (see Fig.~\ref{fig:beta}) clearly shows the different subsystems making up Abell 2256. 
The ``mouth'' appears to be an infalling system, which is, in projection, approximately 250 kpc away from the main core, as we will discuss in Sect.~\ref{sect:discussion_CF}. To the south and east, the beta-model subtracted image in Fig.~\ref{fig:beta} reveals a large tail-like structure of a projected length of $\sim$600 kpc. This feature is likely associated with the other cooler subcluster, the ``shoulder'' (Sect.~\ref{sect:results_morphologies}), which appears to be much larger than previously reported by \citet{Sun02} based on shallower {\it Chandra} observations.

\begin{figure*}
	\includegraphics[width=\textwidth]{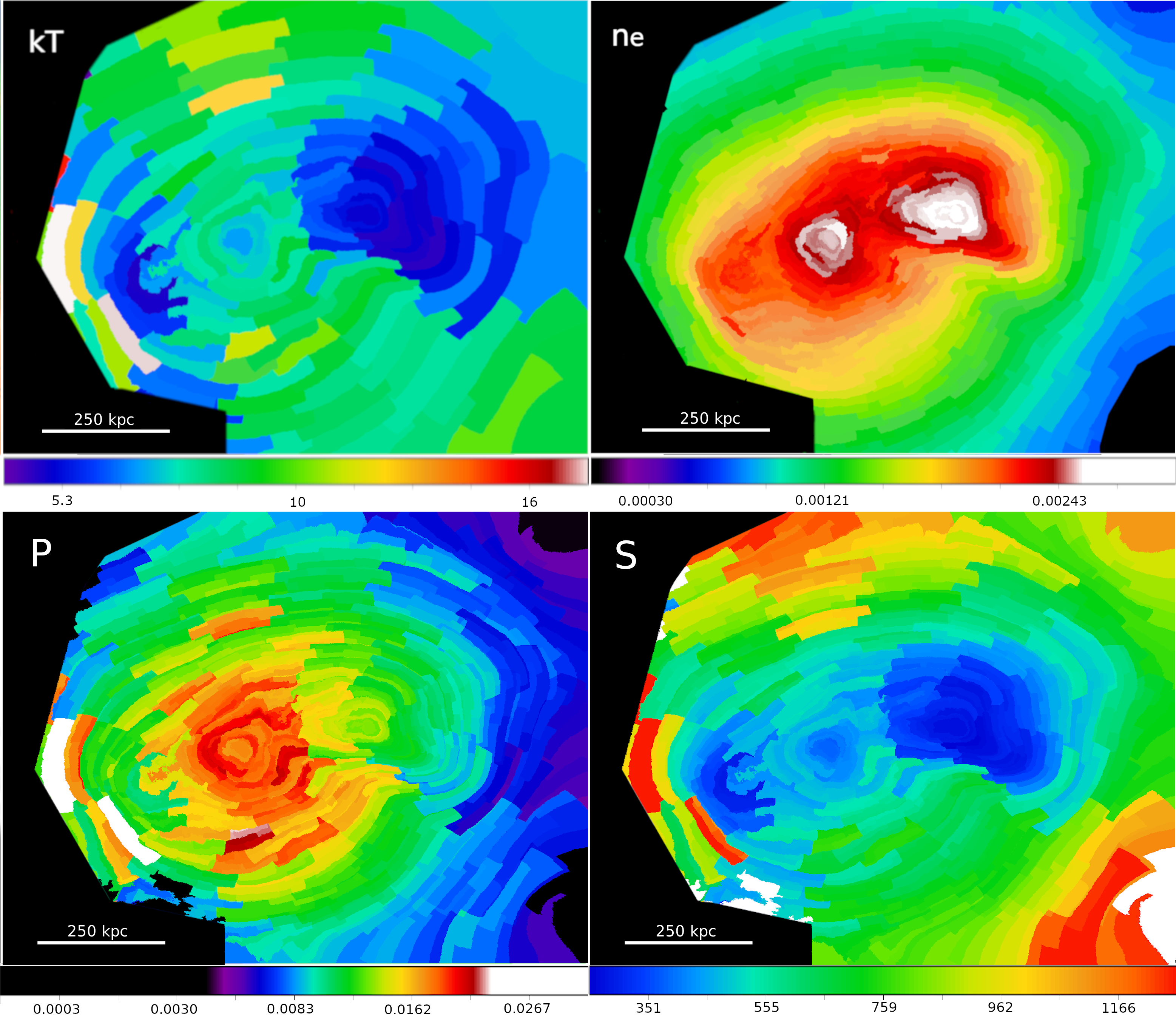}
	\caption{Thermodynamic maps of temperature (\textit{top left}), electron density (\textit{top right}), pressure (\textit{bottom left}), and entropy (\textit{bottom right}) with corresponding units of keV, cm$^{-3}\times$($l/1$Mpc)$^{-1/2}$, keVcm$^{-3}\times$($l/1$Mpc)$^{-1/2}$, and keVcm$^2\times $ ($l/1$ Mpc)$^{1/3}$. The temperature and electron density maps are produced where each region has a S/N of 70 (5,000 counts) and 33 (1,000 counts) respectively; while the pressure and entropy maps are produced using the results from the top two different S/N maps. In the temperature map, it is clear that the two substructures seen in Fig.~\ref{fig:beta} are actually cold fronts produced from the merging subclusters. The center of the main cluster component appears to also be relatively cool, while the temperature increases radially to the north and south of the center. The electron density map shows that the infalling subcluster has a larger density than the main cluster component, but also that the distribution is far from symmetrical due to the previous merging event. The pressure map shows that the main cluster component also contains the main pressure peak, also worth noting are regions showing significant pressure jumps to the north and south of the cluster. These regions to the north and south also show increased entropy, indicating that they have had some previous involvement in the merger history.}
	\label{fig:jointmaps_black}
\end{figure*}

\subsubsection{The ``Shoulder'' Tail}
\begin{figure*}
        \includegraphics[width=\textwidth]{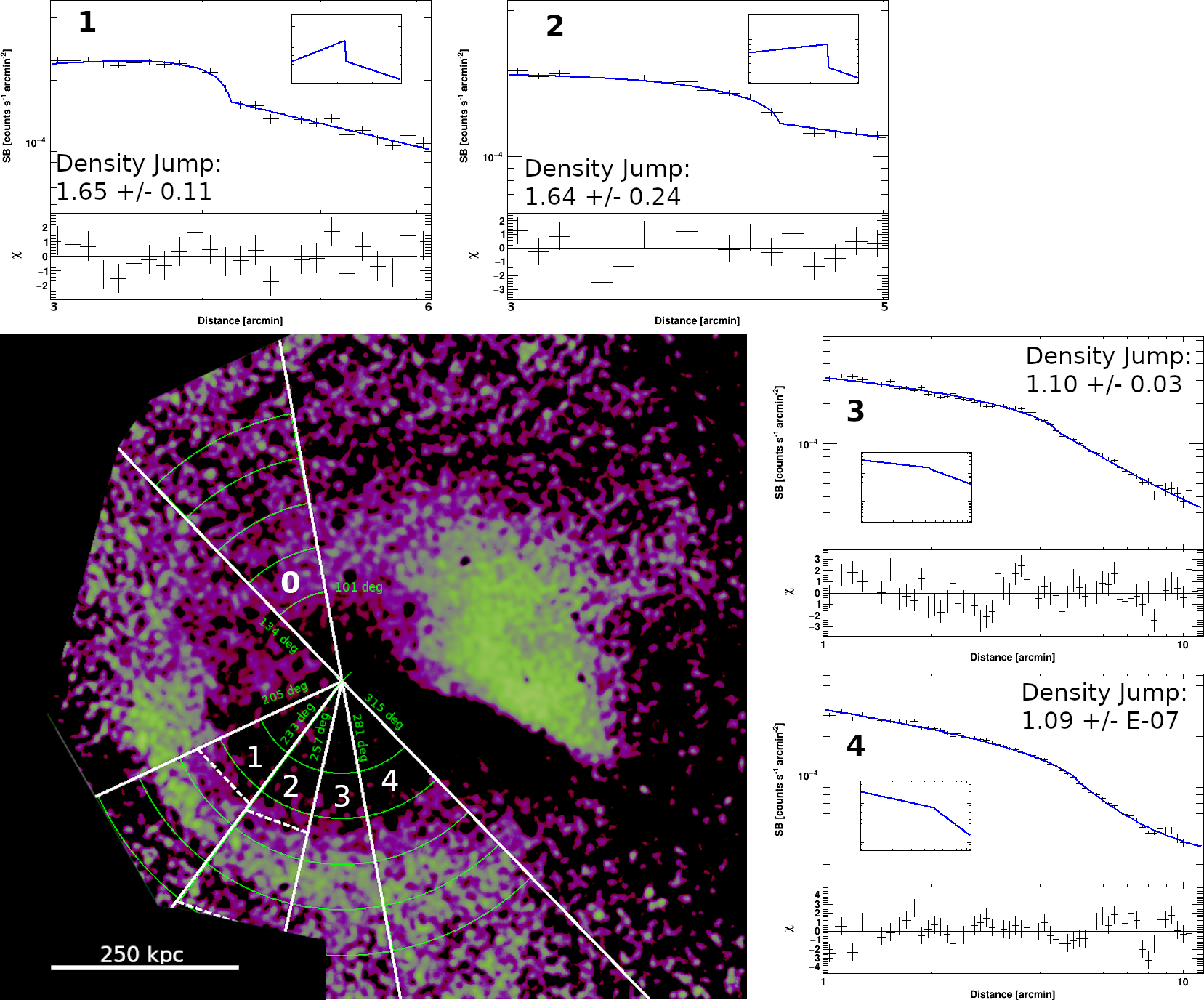}
	\caption{\textit{Center}: The beta model residual of Abell 2256, solid white lines indicate the wedges taken for the surface brightness profiles, originating at the best fit center determined via Monte Carlo methods, dashed white lines indicate starting point for surface brightness profiles.  North wedge labeled 0 corresponds to a region free of emissions from the ``shoulder'' and ``tail'', used in determining a mass estimate for the excess gas in the ``tail''. Green radial sectors in each wedge are used in calculating the profile seen in Fig.~\ref{fig:tail-tempdens}. White numbers in the wedges correspond to the surface brightness plots of the same number.} \textit{Top-Clockwise}: Zoomed in surface brightness profiles across the ``tail'' region showing where a broken power law fit density jumps occur. Clockwise plots corresponds to left to right wedges in the south of the image.
	\label{fig:tail}
\end{figure*}

The surface brightness discontinuity associated with the ``shoulder'' indicates the presence of a density jump of $\sim1.7$.  Fig.~\ref{fig:tail} shows surface brightness profiles across the new tail feature.  
The surface brightness wedges across the tail closest to the ``shoulder'' have density jumps of 1.65, consistent with the ``shoulder'' itself, while the surface brightness wedges across the southern and southwestern part of the tail show more modest jumps. 

Interesting to note, the radio halo observed at low frequencies appears to have a `step' in the radio flux which is roughly cospatial with the X-ray tail feature. This `step' has been noticed in previous work \citep[see Fig. 3 in][]{Tracy2011}, before the tail-like X-ray feature was observed, so no explicit connection could be made to X-ray data.

\begin{figure}
	\includegraphics[width=\columnwidth]{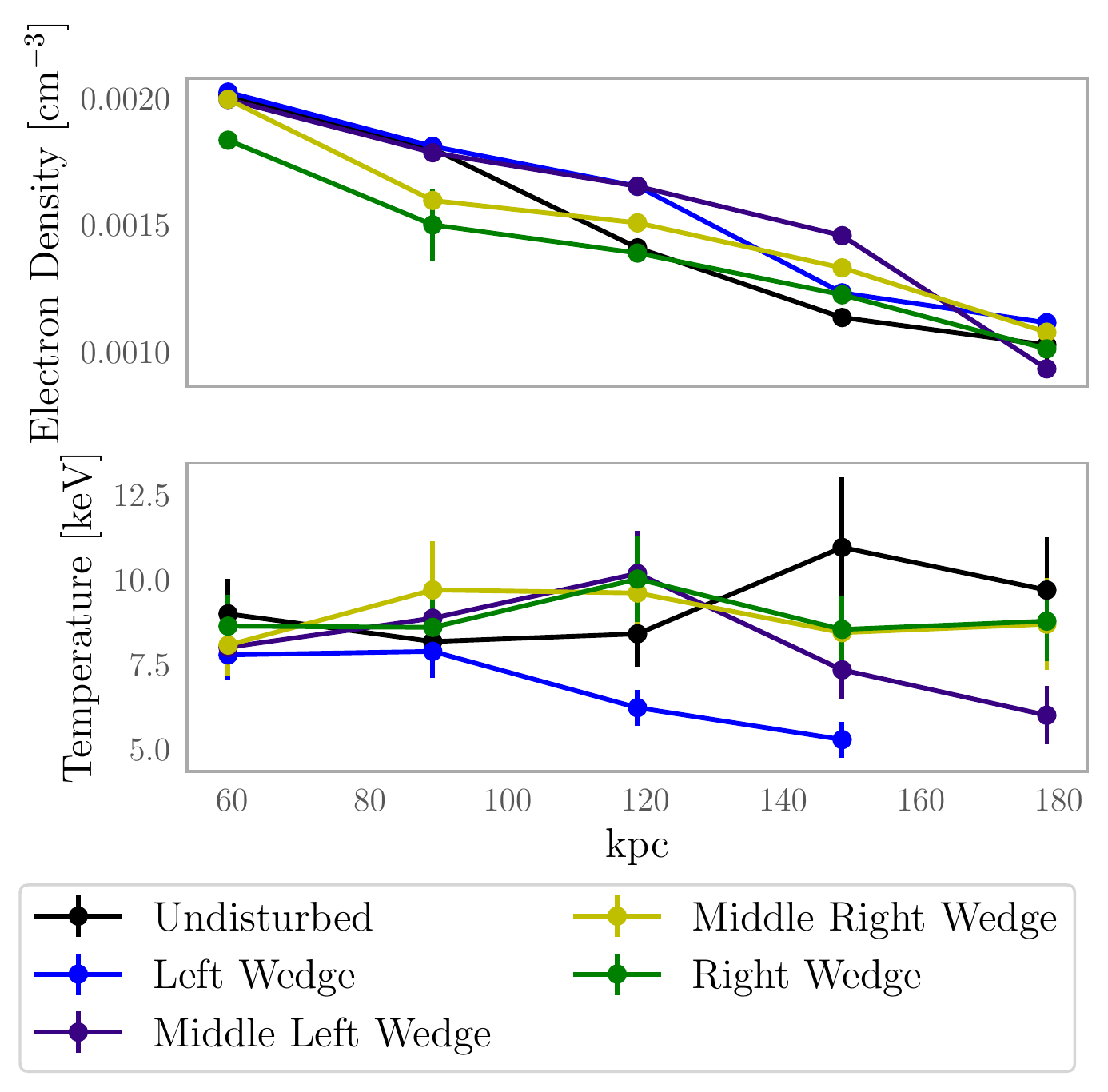}
	\caption{Projected radial profiles of the gas density and temperature from the five different wedges shown in Fig.~\ref{fig:tail}. The last temperature value from the left wedge was unconstrained and removed. Temperatures are consistent with being mostly constant across the tail, despite the increased electron density.}
	\label{fig:tail-tempdens}
\end{figure}

Fig.~\ref{fig:tail-tempdens} shows the projected gas density and temperature profiles for all of the wedges across the bright tail; as it contains no excess gas features, we use the northern wedge shown in Fig.~\ref{fig:tail} as a proxy for an `undisturbed' profile, plotted as a black line. Comparing the wedge profiles with the undisturbed profile shows a density excess associated with the apparent tail. However, we do not see a significant temperature change across this feature.

\subsubsection{The ``Mouth'' and the ``Arc''}

\begin{figure*}
        \includegraphics[width=\textwidth]{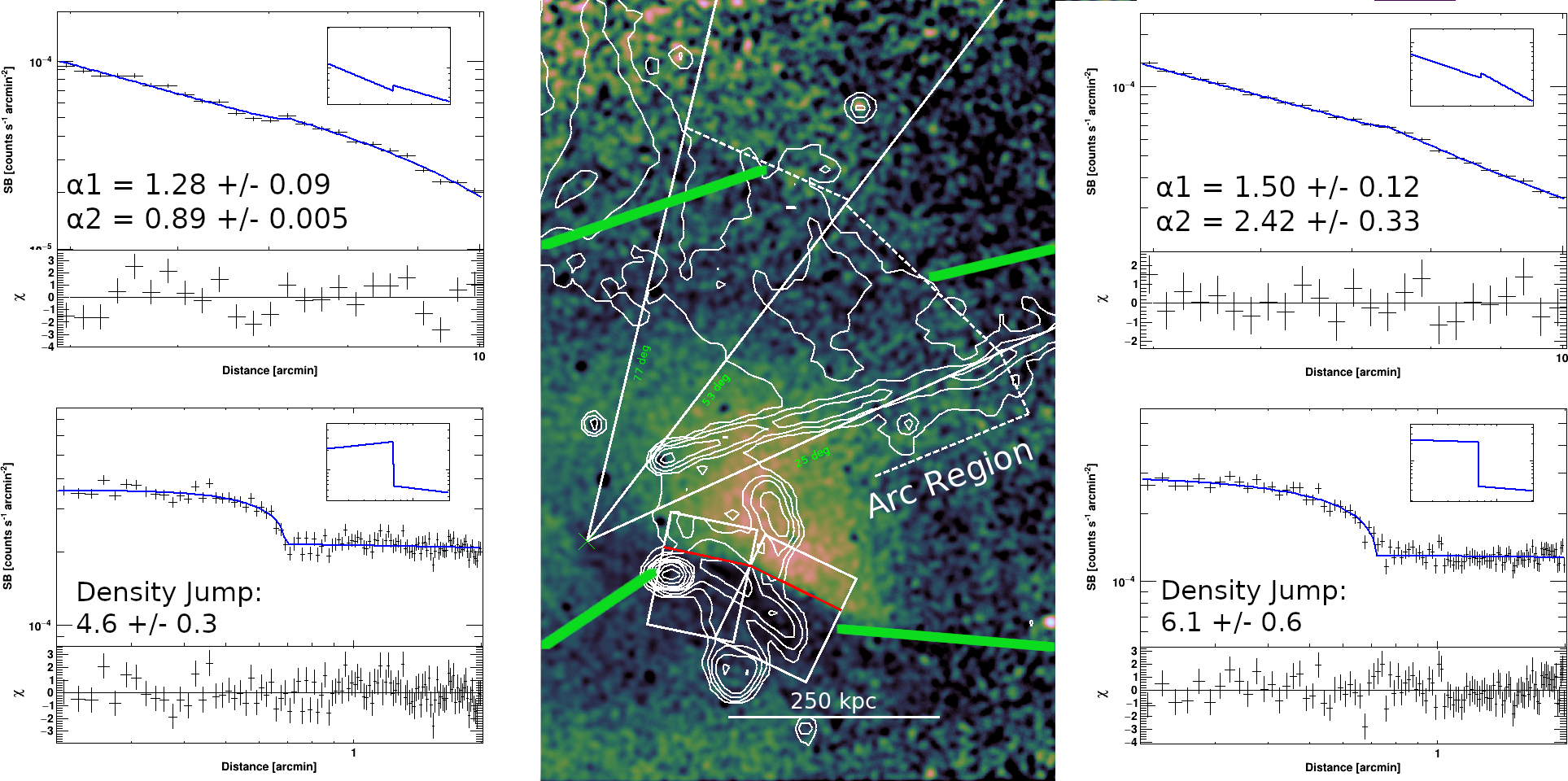}
	\caption{\textit{Center}: The beta model residual of Abell 2256, zoomed in on the bright ``mouth'' and faint arc region, with additional contours from the GMRT 325 MHz radio image. White lines indicate the wedges taken for the surface brightness profiles, originating at the best fit center determined via Monte Carlo methods. Dotted lines indicate the location of the arc as determined by the broken power law fits. Green lines indicate which plots correspond to which regions. Red lines indicate the ``mouth'' cold front. \textit{Top}: Zoomed in surface brightness profiles across the arc region showing where a broken power law fit density jumps of $\sim1.14\pm 0.04$ and $\sim1.10\pm 0.02$, left and right plots are of the left and right wedges, respectively. \textit{Bottom}: Surface brightness profiles across the left and right regions of the ``mouth'' cold front where broken power laws fit density jumps of $\sim4.6\pm 0.3$ on the left and $\sim6.1\pm 0.6$ on the right.} 
	\label{fig:cflr}%
\end{figure*}

To investigate the prominent discontinuity at the mouth, we derive two surface brightness profiles: one across the eastern and another across the western part of the cold front. The profiles are shown in Fig.~\ref{fig:cflr} and are derived using $\sim100$~kpc long box regions parallel to the discontinuity, binned by 1.2~kpc (1 arcsec). We fit the surface brightness profile under the assumption that the corresponding 3D gas density distribution can be described by a broken power-law. Our best fit model shown in Fig.~\ref{fig:cflr} is consistent with a discontinuity that is narrower than the image resolution of 1.2 kpc and indicates a presence of a density jump of a factor of $\sim$4.6$\pm 0.3$ in the eastern part of the front, and $\sim$6.1$\pm 0.6$ in the western part of the front.

The model divided residual image also reveals a surface brightness excess, labeled the ``arc'' in Fig.~\ref{fig:beta}, which corresponds to the outer edge of the radio relic. We confirm the presence of the arc by extracting surface brightness profiles. As shown in Fig.~\ref{fig:cflr}, these profiles can be best described with a broken power-law density distribution, with an inner slope of 1.5$\pm 0.1$, outer slope of 2.4$\pm 0.3$, and a density increase of 1.14$\pm 0.04$ in the western part, and with an inner slope of 1.3$\pm 0.1$, outer slope of 0.90$\pm 0.01$, and a density increase of 1.10$\pm 0.02$ in the eastern part of the relic.



\subsection{Spectral Analysis}

The 2D temperature map in Fig.~\ref{fig:jointmaps_black} (top left panel) shows that the main cluster core appears to be relatively hot, with an average temperature around $kT\sim7$ keV. 
To the north and southwest of the main cluster, the temperature map reveals regions with $kT$ above 10~keV. The ``mouth'' and ``shoulder'' are associated with cooler ($\sim$4 keV) gas. 


The pressure map shows that the main pressure peak is near the center of the main cluster component. The overall pressure distribution is highly asymmetric, with elongated features extending mostly to the north and southwest. 
The entropy map reveals that the infalling subclusters are associated with lower entropy gas. The main cluster component is also of relative low entropy. It is important to note that the assumption of a constant 1 Mpc line of sight distance in the thermodynamic maps implies that the density and pressure are potentially underestimated and the entropy is overestimated in the main core and the two subclusters.


About 180 kpc (3.3 arcmin) 
to the north and 130 kpc (2.5 arcmin) to the south from the main pressure center of the cluster, 
are regions indicating possible temperature jumps and increased pressure. 
These features are discussed in detail in Section~\ref{sec:shocks}. 




\section{Discussion}\label{sec:discussion}

\subsection{The Cold Front at the ``Mouth''}\label{sect:discussion_CF}

\begin{figure*}
	\includegraphics[width=\textwidth]{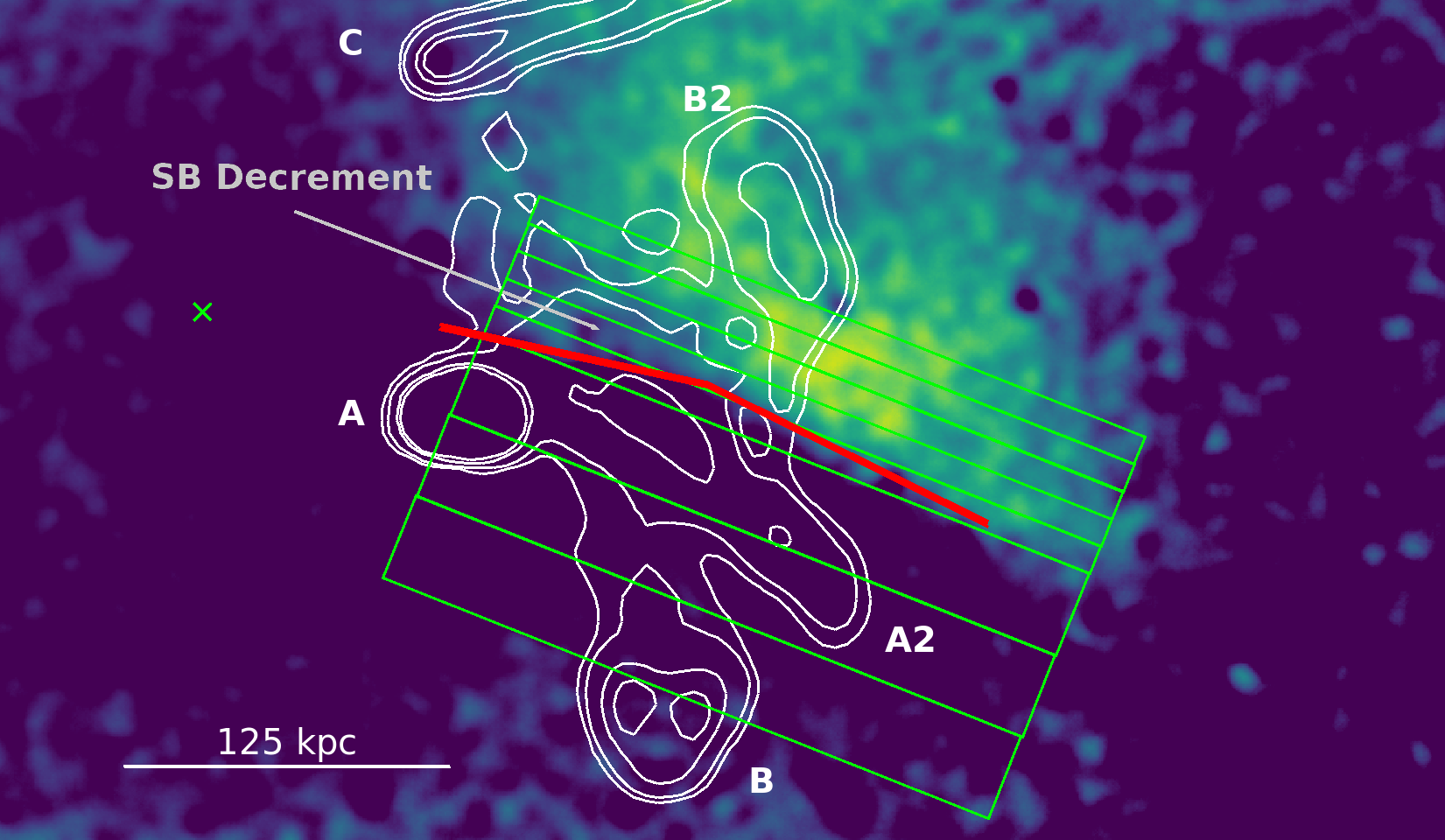}	
	\caption{Beta model residual of the X-ray image of Abell 2256 with the 325 MHz GMRT radio contours overplotted in white (these contours have the same spatial resolution, but now show fewer contour levels). The surface brightness edge at the ``mouth'' is indicated with a red line. The green boxes indicate the specific regions chosen to probe the general profile across the cold front. The radio galaxy labeled `A' is the BCG of an infalling group and it is displaced $\sim50$~kpc to the south of the cold front. The displacement of the galaxy could be the result of ram-pressure stripping during the infall. The low frequency radio plasma originating in the BCG (and possibly partly in the radio galaxy labeled `B') appears to be interacting both with the low entropy gas on the bright side of the surface brightness discontinuity (see source B2) and with the hot gas on the faint side of the cold front (see source A2).  
	}
	\label{fig:coldfront}
\end{figure*}

The most striking feature in the X-ray image of Abell 2256 is the prominent 250~kpc long cold front at the southern edge of the bright, low entropy gas cloud that we labeled ``mouth''. A careful analysis of {\it Suzaku} X-ray spectra by  \citet{Tamura11} indicates that this gas cloud is infalling with a radial velocity of 1500~km~s$^{-1}$. However, it is not associated with any bright galaxies. By analysing the optical data of the cluster, \citet{Berrington02} identify two subgroups: the main cluster, with its BCG at the position of the central surface brightness excess (the ``body'') and an infalling group with its brightest galaxy associated with the radio galaxy `A', which can be found $\sim 50$ kpc south of the eastern edge of the cold front in projection (see Fig.~\ref{fig:coldfront}). The fact that this BCG is displaced with respect to the infalling low entropy gas strongly suggests that the hot gas of this infalling group has been stripped during its merger with the main cluster. Assuming that the dark matter follows the galaxy distribution, we predict that it is also significantly offset from the infalling low-entropy gas.

The prominent 250~kpc long cold front shows a break and a possible surface brightness decrement on its bright side, which overlaps with low frequency radio emission. 
In Fig.~\ref{fig:coldfront}, we show the {\it Chandra} X-ray image with the overplotted 325~MHz GMRT radio contours. The image suggests that the radio plasma might be interacting with the low entropy X-ray emitting gas associated with the cold front. The eastern side of the cold front overlaps with the radio contours. The radio emission then extends 150~kpc in projection further to the northwest, forming the source labeled `B2'. Interestingly, B2 coincides with the eastern edge of a surface brightness depression in the X-ray image. The southern part of the radio tail of the galaxy extends 200~kpc in projection to the southwest. 

\citet{Intema09} proposed that the low-frequency radio emission labeled `B2' in Fig.~\ref{fig:coldfront} originates from the radio galaxy labeled `B', which, based on its redshift, is not a member of an infalling group that could be associated with the low-entropy gas.
However, both the break and the X-ray surface brightness decrement in the cold front that separates the infalling gas from the ambient cluster appear to be associated with the radio plasma. This potential association indicates that the radio plasma most likely originates from and has been left behind by the displaced BCG of the infalling group (source A). The radio bright regions appear to be associated with X-ray surface-brightness decrements indicating that the radio plasma could form cavities and/or provide significant additional non-thermal pressure support. As originally proposed by \citet{Intema09}, the source A2 also most likely originates in the source A and the radio plasma is stripped by the ordered southwestward motion of the ambient ICM around the cold front. 

The radio emission labeled A2 and B2 in Fig. \ref{fig:coldfront} is remarkably steep and it is not visible at frequencies above 1 GHz \citep[for a comparison, see the 1 GHz VLA image from][]{Owen14}.
The aging radio plasma outside the cold front (source A2) could have been revived, when the differential motions around the cold front amplified the local magnetic fields and thus also the synchrotron emission at low frequencies. A testable prediction of this scenario is that the low frequency radio emission at the contact discontinuity is polarized. The radio plasma on the bright side of the discontinuity could also have been revived by the turbulent motions expected inside of an infalling low entropy cloud that is being stripped. 

While on large scales, we observe a break and an surface brightness decrement, on small scales the cold front is remarkably sharp, indicating a strong suppression of transport processes across the discontinuity. Surface brightness profiles shown in Fig.~\ref{fig:cflr} indicate that the discontinuity is narrower than $\sim3$~kpc.

To investigate how the temperature and pressure profiles behave across the cold front, we analyse spectra extracted from regions that are 250 kpc (220 arcseconds) wide and 11 or 34 kpc (10 and 30 arcseconds, respectively) thick, shown in Fig.~\ref{fig:coldfront}. The thinner regions, extracted from the brighter and cooler side of the cold front, contain $\sim6000$ counts and the thicker regions, extracted from the fainter and hotter side of the contact discontinuity, contain around 10000 counts. 
Fitting the spectra across the ``mouth'' using a single temperature model with photoelectric absorption, \texttt{phabs(apec)}, shows that the temperature increases from 4.8$\pm 0.2$~keV to 7.9$\pm 0.4$~keV. Deprojecting the bright low-entropy component by fitting the spectra using an absorbed two-temperature model, \texttt{phabs(apec+apec)}, and tying the temperature and the normalization of one of the apec models to the best fit values of the hotter gas outside the front, we obtain a temperature of 2.2 $\pm 0.5$~keV. This corresponds to a factor of 3.6$\pm 0.8$ temperature jump from the cold to the hot region, smaller than the factor of 4.6$\pm 0.3$ and 6.1$\pm 0.6$ density jumps determined by fitting the surface brightness profiles. \citet{Sun02} observed a relatively high iron abundance ($\sim 0.6$ solar) within the inner 150 kpc of the ``mouth''. We similarly observe an above average level of iron enrichment inside the ``mouth'', with values between $\sim 0.4$ and $\sim 0.6$, which abruptly drop to values between $\sim 0.1$ and $\sim 0.2$ when crossing to the faint outside regions. We note that these observed features in the abundance profiles are consistent with known literature of sloshing-type cold fronts, which have been shown to help redistribute metals already present in the ICM \citep{Mernier2018}.

We note that the inferred density jumps from the broken power law fits exceed the temperature jump.  We speculate that the assumption of spherical symmetry, used in obtaining the jumps, might not hold due to the pile up of gas around the cold front. Furthermore, the systematic uncertainties associated with the deprojected temperature of the infalling gas could be significantly larger than the cited statistical uncertainties. Due to these uncertainties, we caution against the over-interpretation of the fit results.  However, this discrepancy may indicate that the infalling dense low entropy gas is over-pressured and expanding as it is stripped from the underlying dark matter halo. This interpretation is supported by the displacement of the BCG from the gas with which it was associated. The offset of the dark matter distribution from the X-ray gas is a clear prediction of our interpretation. While, to our knowledge, no weak lensing analysis has been published for this cluster, future observations could confirm or rule out our proposed scenario.

The Coulomb mean free path of the electrons outside the ``mouth'' is about 20~kpc. The width of the discontinuity determined from the surface brightness profiles is unresolved and the distance within which we measure the temperature increases from 5~keV to 8~keV is $\sim20$~kpc. If the transport processes were not suppressed, the expected width of the front would be several mean free paths \citep[for suppression of transport processes across other cold fronts see also e.g.][]{ettori2000,vikhlinin2001,Maxim07,ichinohe2015,werner2016a}. 


Shear amplification of magnetic field lines parallel to a cold front surface (potentially associated with the low-frequency radio emission source `A2') could suppress both small-scale Kelvin-Helmholtz instabilities and prevent conduction/diffusion across the front \citep{Zuhone13}. Magneto-hydrodynamic simulations show that magnetic fields can have similar effect as viscosity on the ICM, inhibiting the mixing of the hotter and cooler flows. 

\subsection{The Proposed Merger Scenario}
There have been many different merger scenarios proposed in the literature. 
As discussed in the previous sections, it is quite clear that Abell 2256 hosts at least three separate infalling subclusters in different stages of a merger, two of which are clearly visible in the X-ray data. 
Previous results from \citet{Sun02}, which introduced the ``shoulder'', were limited by a shallow observation and could only identify the approximate dense core of the subsystem. The approximate length of the tail of the ``shoulder'' cold front appears to be around 600 kpc in projection, much larger than previously reported, lending evidence that the timescale for that particular merging event is much longer than that of the subcluster at the western ``mouth'' in the pre-merger phase. 
Based on the position of the eastern cold front associated with the ``shoulder'', the merging event most likely occurred off-center and could be responsible for the large low frequency radio halo reported in \citet{Tracy06}. The long tail is most likely associated with gas stripped from the infalling group associated with the ``shoulder''. Interestingly, however, the temperature distribution across this feature does not show a decrease expected for stripped low entropy gas (see Fig.~\ref{fig:tail-tempdens}). This indicates that the stripped gas is being heated by mixing or conduction, possibly by shocks associated with the merger of the ``mouth'' which occurs after the formation of the tail. A detailed study of this feature will be published elsewhere.

The subcluster associated with the low entropy ``mouth'' fell in later and appears in projection $\sim250$~kpc before core passage. This subcluster appears to move through the ambient ICM supersonically, with a large line-of-sight velocity component \citep[see][]{Tamura11}. Even though in projection, the merger appears to be pre-core passage, along our line-of-sight the subcluster could already be well past the midplane of the main cluster. 

The temperature and pressure maps in Fig.~\ref{fig:jointmaps_black} show indications for possible shocks, which are discussed in Sec.~\ref{sec:shocks}.  

\subsection{The Candidate Shock Regions}\label{sec:shocks}
The regions of high temperature and pressure to the north and south of the pressure peak seen in Fig.~\ref{fig:jointmaps_black} might correspond to potential shocks. Using a photoelectrically absorbed single temperature model to examine the regions, would, according to the Rankine-Hugoniot jump condition for temperature jumps, correspond to shocks with a Mach number of $\sim1.6$. Candidate regions pertaining to the shock, as well as regions preceding and proceeding the front, were additionally examined using a photoelectrically absorbed two-temperature model, with one free temperature component and another frozen to 0.2 keV, to account for possible differences between the subtracted blank sky observations and the actual foreground Galactic halo emission. However, the 0.2 keV component was not significant. Furthermore, we varied the background scaling parameter to investigate how it affects our results. An increase or decrease of the scaling parameter by 10\% did not change the conclusions. 

The high-temperature regions are, however, not associated with any corresponding features in the surface brightness distribution and they are not equidistant from the main pressure center of the cluster (they are seen 225 kpc to the north and 170 kpc to the south). This could possibly be due to a complex geometry, but has also traditionally been difficult to test through observational techniques. While literature studies usually attribute differences in the shock properties measured using the X-ray surface brightness and spectroscopic temperature to the complex geometry, limited quantitative analysis on these effects is available \citep[eg. Fig. 11 in][]{Akamatsu2017}.


To get a simpler view on the role of projection effects in determining the observed X-ray surface brightness, we generate a purely geometric toy model of an inclined shock. We consider a merging system with both shock and cold fronts with a hyperbolic geometry,  similar to the one employed in \citet{Luca2019}. To account for the azimuthal variation of the Mach number along the shock front, we consider the jump conditions for oblique shock waves as in Chapter 92 of \citet{Landau1987Fluid}. Specifically, the density enhancement at each position of the shock can be written as,
\begin{equation}
    \frac{\rho_2 - \rho_1}{\rho_1} = \frac{2(\mathcal{M}_1^2\sin^2\phi-1)}{(\gamma -1)\mathcal{M}_1^2\sin^2\phi +2}.
\end{equation}
where $\phi$ is the angle between the tangent to a given point of the shock surface and the axis of symmetry of the shock front. We also look at different Mach numbers for the shock ($\mathcal{M} = $1.5, 2.0, and 3.0), all of these toy model figures are included in the online supplement to this paper. A similar description but with an inverse direction in the temperature jump is considered for modelling the cold front. This toy model additionally considers: 1) instantaneous electron heating 2) the standoff distance is not set to have a physical meaning \citep[as in][]{zhang2019}, but adapted to the observed cluster morphology 3) as it is purely geometrical, the model does not account for any perturbations in the thermodynamics of the ICM. Due to some of these limitations, more detailed simulations are considered.

\begin{figure}
\centering
	\includegraphics[width=\columnwidth]{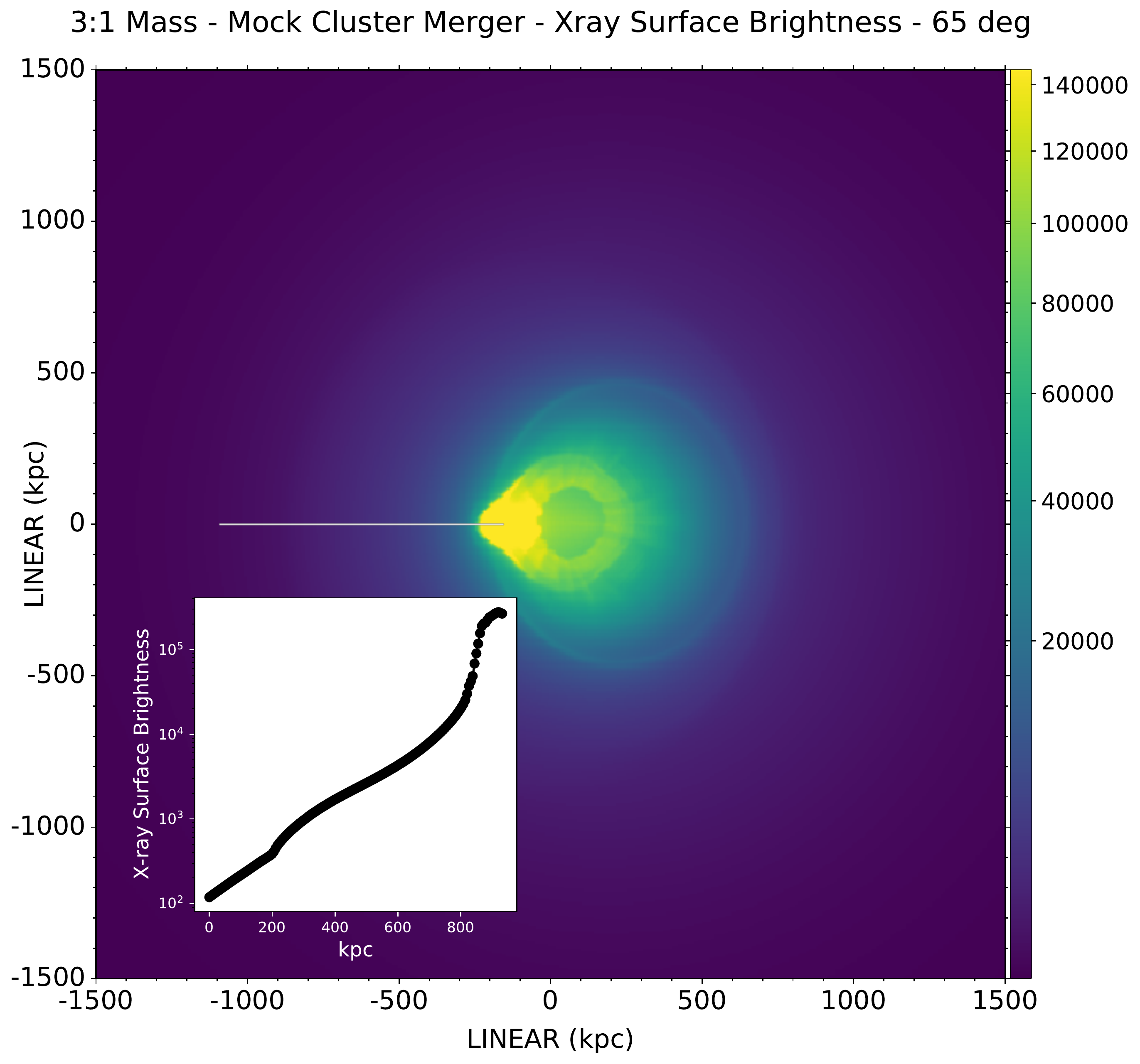}	
	\caption{Simulated X-ray surface brightness image of a galaxy cluster merger with a mass ratios of 3:1 at a 65 degree viewing angle (where 90 degrees is along the merger axis, and 0 degrees is perpendicular to the merger plane). The white line indicates the cross section in which the inlaid plot quantifies the surface brightness profile across the front.}
	\label{fig:shockmodel3-1_65sb}
\end{figure}

\begin{figure}
\centering
	\includegraphics[width=\columnwidth]{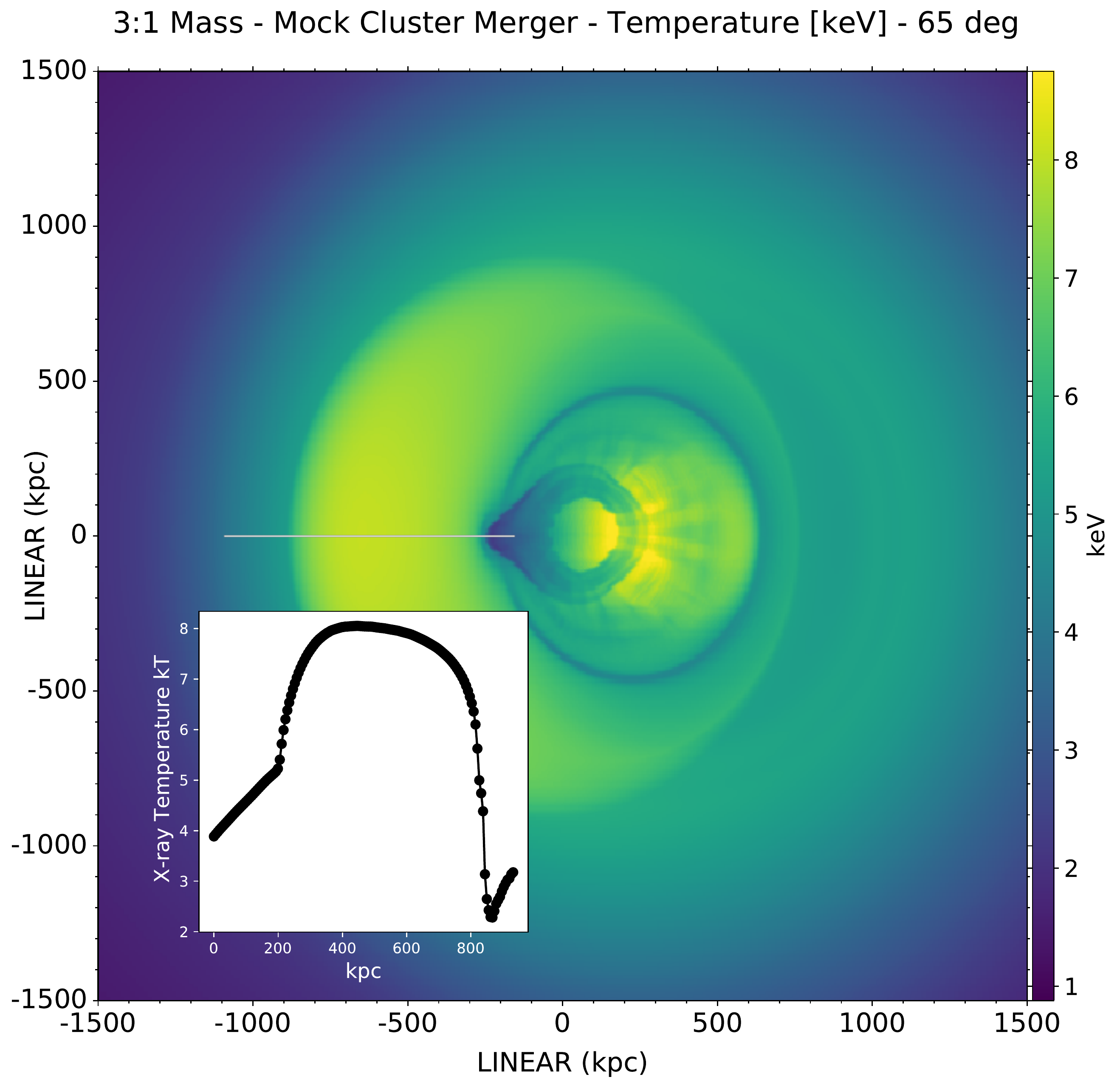}	
	\caption{The same as Fig. \ref{fig:shockmodel3-1_65sb}, but showing the simulated X-ray temperature distribution.}
	\label{fig:shockmodel3-1_65kt}
\end{figure}

\begin{figure}
\centering
	\includegraphics[width=\columnwidth]{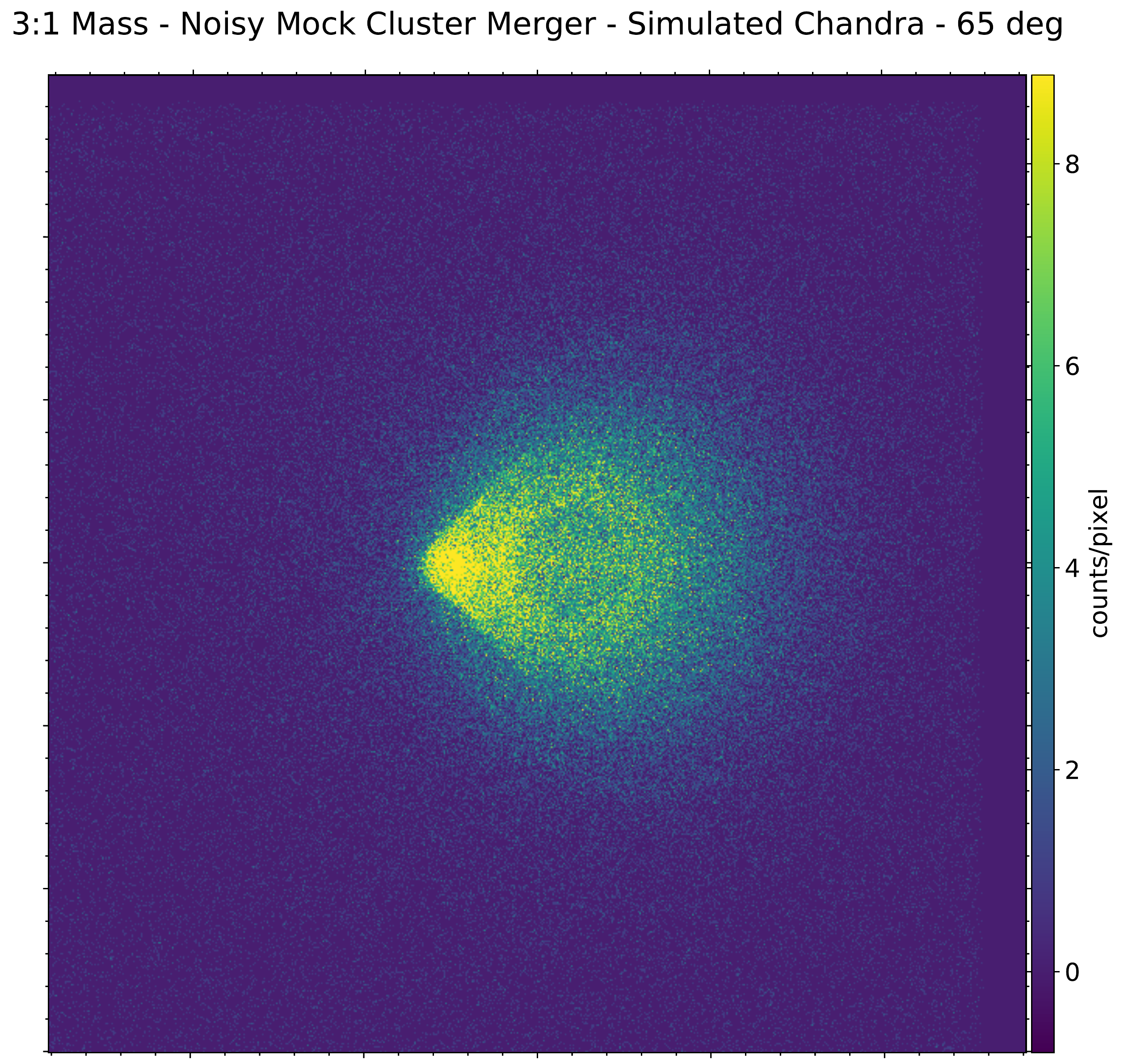}	
	\caption{A mock {\it Chandra} X-ray image of the simulation in Fig. \ref{fig:shockmodel3-1_65sb}, using the ACIS-I response for Cycle 22 with an exposure time of 200 ks, assuming the redshift of Abell 2256, including astrophysical foreground and absorption, as well as the ACIS-I instrumental background.}
	\label{fig:shockmodel3-1_65noise}
\end{figure}

To further test this, we examined various galaxy cluster mergers with zero impact parameter collisions and different mass ratios (1:2, 1:3, and 1:5). These simulations were performed using the \code{FLASH} \citep{Dubey2009} and \code{GAMER} \citep{Schive2018} AMR codes as applied in previous work \citep{ZuHone2011}. The simulation data were projected to produce maps of X-ray surface brightness and ``spectroscopic-like'' \citep{Mazzotta2004} temperature along lines of sight in the range of 0 to 90 degrees, where 0 degrees is perpendicular to the merger plane, and 90 degrees is along the merger axis. Counts images were also simulated using the \texttt{pyXSIM} and \texttt{SOXS} packages \citep{ZuHone2014,ZuHone2016}, using the \textit{Chandra} ACIS-I responses for Cycle 22 with an exposure time of 200 ks at the redshift of Abell 2256. These mocks also include astrophysical foreground and absorption, as well as the ACIS-I instrumental background. Both of these processes were also performed for the toy model case. Images of all of our mock observations are available in the online supplementary document. \citet{Berrington02} estimated a 3:1 mass ratio for Abell 2256 using optical data, therefore a similar mass ratio was used to create Fig.~\ref{fig:shockmodel3-1_65sb}, Fig.~\ref{fig:shockmodel3-1_65kt}, and Fig.~\ref{fig:shockmodel3-1_65noise}, which show the X-ray surface brightness, the temperature, and a mock {\it Chandra} ACIS-I image, respectively.

Fig.~\ref{fig:shockmodel3-1_65sb} and Fig.~\ref{fig:shockmodel3-1_65kt} show that, when seen near face-on at angles close to our line of sight, temperature jumps are possible to observe without sharp associated surface brightness features and density discontinuities. When noise is added to the simulation results like in real observations (see Fig.~\ref{fig:shockmodel3-1_65noise}), small scale surface brightness features disappear almost entirely, leaving only temperature (and pressure) jumps as indicators that a shock front is present.


The prominent radio relics also indicate the presence of shocks propagating in the cluster outskirts. 
\citet{Trasatti15} looked at the overall spectrum of the relics, where the steepening between the frequency bands of 63-1369 MHz and 1369-10450 MHz 
translates to approximate radio Mach numbers of 2.6 and 2.2, greater than the values determined from the X-ray temperatures for the shock candidates located at smaller radii. Unfortunately, the X-ray emission is too faint to perform a detailed analysis at the radii where the relics are seen. However, the surface brightness distribution reveals an X-ray faint arc that appears to be confining the relics at large radii (see Fig. \ref{fig:cflr}). The nature of this arc is not clear and deserves a further detailed analysis using deeper X-ray observations. Overall the multiple inner candidate shock fronts with a prominent radio relic located further out in radius are somewhat reminiscent of the thermodynamic structure of CIZA J2242.8+5301, another well studied merging cluster \citep[see Figure 8 of][]{Ogrean14}.

\section{Summary}
For the last 50 years, Abell 2256 has been an exciting and somewhat mysterious galaxy cluster due to the exceptional substructures and rich variety of radio morphologies. In this paper, we presented results from deep X-ray observations of the merging cluster Abell 2256 to try to understand the merger history in an effort to reconcile the available multi-wavelength observations. To summarize the results:
\begin{itemize}
    \item Deep {\it Chandra} and {\it XMM-Newton} X-ray imaging and spatially-resolved spectroscopy of Abell 2256 reveals three subclusters: (i) the main cluster ``body'', (ii) the ``shoulder'', which is a remnant of an older merger in the eastern part of the cluster with a $\sim600$ kpc long tail, and (iii) the ``mouth'', which is a bright, bullet-like, infalling low-entropy system, with a large line-of-sight velocity component.
    
    \item The low-entropy system displays a 250 kpc long cold front with a break and an intriguing surface brightness decrement overlapping with radio emission.
    
    \item  The radio loud BCG associated with the infalling group appears dissociated from the infalling low-entropy gas. We predict that the dark matter is also offset from the stripped gas. 

    \item The ``mouth'' cold front appears to be interacting with the low-frequency radio emitting plasma potentially associated with the BCG. Some of the observed radio emission might be revived by magnetic field amplification due to differential gas motions, and we predict might thus be strongly polarized.
    
    \item The strikingly smooth and sharp front at the ``mouth'' is narrower than the Coulomb mean free path of the ambient gas, indicating suppression of transport processes, possibly by the amplified magnetic fields.
    
    \item We observe a low frequency radio surface brightness `step' coincident with the long tail-like X-ray feature.

    \item Based on our projected thermodynamic maps, we discuss the possibility that the infall of the subcluster is generating shocks. 
    
    \item Various cluster merger simulations and shock models were used to test the effects of different geometries and lines-of-sight on the observable results (see the online supplementary material).
    
    \item The high temperature potentially shocked gas and location of the southern extent to the radio relic support the hypothesis that we are observing a large scale shock inclined nearly along the line-of-sight, which would be consistent with the merger scenario presented in \citet{Berrington02}.
\end{itemize}

\section*{Acknowledgements}
The authors would like to thank R. van Weeren for providing excellent GMRT data. The authors would like to thank E. Churazov, C. Sarazin, M. Sun, and D. D. M. Ferreira for helpful discussions. The authors would like to thank the reviewer for the constructive feedback that helped to strengthen the conclusions and improve the quality of the manuscript. JPB acknowledges the support of the studentship at the European Southern Observatory. This work was supported by the Lend\"ulet LP2016-11 grant awarded by the Hungarian Academy of Sciences. Basic research in radio astronomy at the Naval Research Laboratory is supported by 6.1 Base funding.

The scientific results reported in this article are based in part on data obtained from the Chandra Data Archive.
This research has made use of software provided by the Chandra X-ray Center (CXC) in the application packages \textsc{Ciao} and \textsc{Sherpa}; as well as NASA's High Energy Astrophysics Software (HEASoft) packages \textsc{Xspec}. The scientific results reported in this article are also based in part on observations obtained with XMM-Newton, an ESA science mission with instruments and contributions directly funded by ESA Member States and NASA.




\bibliographystyle{mnras}
\bibliography{bib} 






\bsp	
\label{lastpage}
\end{document}